\documentclass[%
 reprint,
 amsmath,amssymb,
 aps,
pra,
]{revtex4-1}

\usepackage{graphicx}% Include figure files
\usepackage{dcolumn}% Align table columns on decimal point
\usepackage{bm}% bold math
\def\beq{\begin{equation}}
\def\eeq{\end{equation}}
\def\bea{\begin{eqnarray}}
\def\eea{\end{eqnarray}}
\def\brcl{\begin{array}{rcl}}
\def\bccl{\begin{array}{ccl}}
\def\blcl{\begin{array}{lcl}}
\def\err{\end{array}}
\def\longeq{\;\;=\;\;}
\def\fatR{{\bf R}}

\def\fatr{{\bf r}}

\begin{document}
%\newcolumntype{S}[2]{D{.}{.}{\#1}}

\title{Atoms in molecules from alchemical perturbation density functional theory}

\author{Guido Falk von Rudorff}
\author{O. Anatole von Lilienfeld}
\email{anatole.vonlilienfeld@unibas.ch}
\affiliation{Institute of Physical Chemistry and National Center for Computational Design and Discovery of Novel Materials (MARVEL), Department of Chemistry, University of Basel, Klingelbergstrasse 80, CH-4056 Basel, Switzerland}

\begin{abstract}
Based on thermodynamic integration we introduce atoms in molecules (AIM) using the orbital-free framework of alchemical perturbation density functional theory (APDFT).
Within APDFT, atomic energies and electron densities in molecules are arbitrary because any arbitrary reference system and integration path can be selected as long as it meets the boundary conditions. 
We choose the uniform electron gas as the most generic reference, and linearly scale up all nuclear charges, situated at any query molecule's atomic coordinates.  
Within the approximations made when calculating one-particle electron densities, this choice affords exact and unambiguous definitions of energies and electron densities of AIMs  
We illustrate the approach for neutral iso-electronic diatomics (CO, N$_2$, BF), various small molecules with different electronic 
hybridisation states of carbon (CH$_4$, C$_2$H$_6$, C$_2$H$_4$, C$_2$H$_2$, HCN), and for all the possible BN doped mutants connecting benzene to borazine (C$_{2n}$B$_{3-n}$N$_{3-n}$H$_6$, $0 \le n \le 3$). 
Analysis of the numerical results obtained suggests that APDFT based AIMs enable meaningful and new interpretations of molecular energies and electron densities. 
\end{abstract}

\maketitle

\section{Introduction}
Methods and approaches to chart chemical compound space (CCS) from an atomistic point of view have been gaining traction for a considerable time~\cite{Beratan1991,ChemicalSpace,ReymondChemicalUniverse,BeratanUnchartedCCS2013,anatole-ijqc2013,mullard2017drug}. 
A quantum mechanics based framework is crucial for exploring CCS in an unbiased way for the purpose of understanding its structure and trends as well as for rational compound design applications~\cite{Beratan1991,Beratan1996,ZungerNature1999,anatole-prl2005,RCD_Yang2006,InverseBeratanYang2008,DesignKeinanBeratanYang2008,fpdesign2014anatole}, as also recently reviewed in the context of catalysis~\cite{batista2019designMLalchemy}.
Unfortunately, atomistically resolved exploration attempts of CCS are severely hampered
due to the electronic Schr\"odinger equation coupling all formally indistinguishable electrons in any molecule or material. Consequently, first principles based bottom-up design efforts can not rely on the virtual build up of optimal materials one atom at a time.

While in principle there are arbitrarily many ways to decompose the electronic energy into atomic contributions, various definitions of energies of atoms in molecules have been proposed, including e.g.~Bader's quantum theory of atoms in molecules ~\cite{bader1975molecular,bader1985atoms,bader1991quantum,popelier1995characterization},
Hirshfeld partitioning~\cite{hirshfeld1977bonded,nalewajski2000information,huang2019density}, or density partition theory~\cite{cohen2007partition}.
It is also possible to infer atomic energies for new out-of-sample molecules from quantum machine learning models~\cite{QMLessayAnatole} trained throughout chemical space, as recently illustrated in 2017~\cite{Amons,DTNN2017}, and subsequently in 2018~\cite{unke2018reactive}.
Here, we propose an alternative, well-defined, efficient, and, in principle, exact approach which provides a unique definition, enabling the calculation of systematic total energy and electron density contributions of AIMs in any arbitrary molecule or material. 

\section{Methods}
\subsection{Theory}
Alchemical perturbation density functional theory (APDFT)~\cite{vonRudorff2018apdft} exploits continuous coupling paths between arbitrary target and reference systems which explicitly include paths without correspondence to reality. Since electronic potential energy and electron density are state functions, meaningful estimates can be obtained by virtue of path integrals. Such computational ``alchemy'' is common in statistical mechanics applications, and can be used, for example, to efficiently estimate free energy changes in drug-binding~\cite{Hansen2014}, solvation of ions~\cite{anatole-jcp2009}, or melting~\cite{sai-iecr2010,sai-pre2011}.
Albeit less frequently studied, exploiting the arbitrariness of the interpolating function as an additional degree of freedom can also be beneficial in the context of {\em ab initio} calculations~\cite{anatole-jcp2009-2,alejandro-jctc2011}.

Assuming a linear interpolation between the initial and final electronic Hamiltonians of any two iso-electronic neutral systems, $\hat{H}(\lambda) = \hat{H}_i + \lambda (\hat{H}_f - \hat{H}_i)$ with 0 $\le \lambda \le 1$, 
thermodynamic integration of the electronic energy over $\lambda$ amounts to the energy difference $E_f - E_i$,
\bea
 \Delta E & = & \int_0^1 d\lambda\, \left.\frac{\partial E}{\partial \lambda}\right|_{\lambda} \longeq  \int_0^1 d\lambda \, \int d\fatr \, \Delta v(\fatr)\, \rho_\lambda(\fatr)\nonumber\\
\eea
where $\partial_\lambda E = \langle \partial_\lambda \hat{H}(\lambda) \rangle = \int d\fatr \, \Delta v(\fatr) \, \rho_\lambda (\fatr)$, according to Hellmann-Feynman theorem~\cite{HF}, and as discussed also in Refs.~\cite{WilsonsDFT,anatole-jcp2009-2,anatole-ijqc2013}. 
Using the chain-rule one can also write, 
\bea
\Delta E & = &  \sum_I \Delta Z_I \int_0^1 d\lambda \left.\frac{\partial E}{\partial Z_I}\right|_\lambda 
\label{eq:alchemy}
\eea
where the sum runs over all those atoms whose nuclear charge $Z$ depends on $\lambda$. In Refs~\cite{WilsonsDFT,anatole-prl2005,anatole-jcp2006-2}, 
$\partial_{Z_I} E =: \mu_I$, is identified as the ``alchemical potential'' of atom $I$, which can be calculated at its position $\fatR_I$ as the electrostatic potential acting on a test-charge exerted by the electron density and all other nuclei but $I$. Its importance for atoms in molecules has already been discussed by Politzer and Murray~\cite{VESPPolitzer1}, and has been evinced for its applicability to computational design efforts of heterogeneous catalysts~\cite{CatalystSheppard2010,DanSheppard-thesis,saravanan2017alchemical,griego2018benchmarking}.
The usefulness of such alchemical first order estimates of changes in properties due to changes in chemical composition has also been explored for covalent bonding~\cite{Samuel-CHIMIA2014}, metals~\cite{MoritzBaben-JCP2016}, semi-conductors with wide direct band-gaps~\cite{Chang2018}, water/BN-doped graphene interactions~\cite{al2017exploring},  or alkali-halide crystals~\cite{AlchemyAlisa_2016}.
Second and higher order alchemical estimates were studied in Refs.~\cite{LesiukHigherOrderAlchemy2012,CCSexploration_balawender2013,Chang2016,balawender2018,Fias2018,vonRudorff2018apdft}.

\begin{figure}
\centering
\includegraphics[width=\columnwidth]{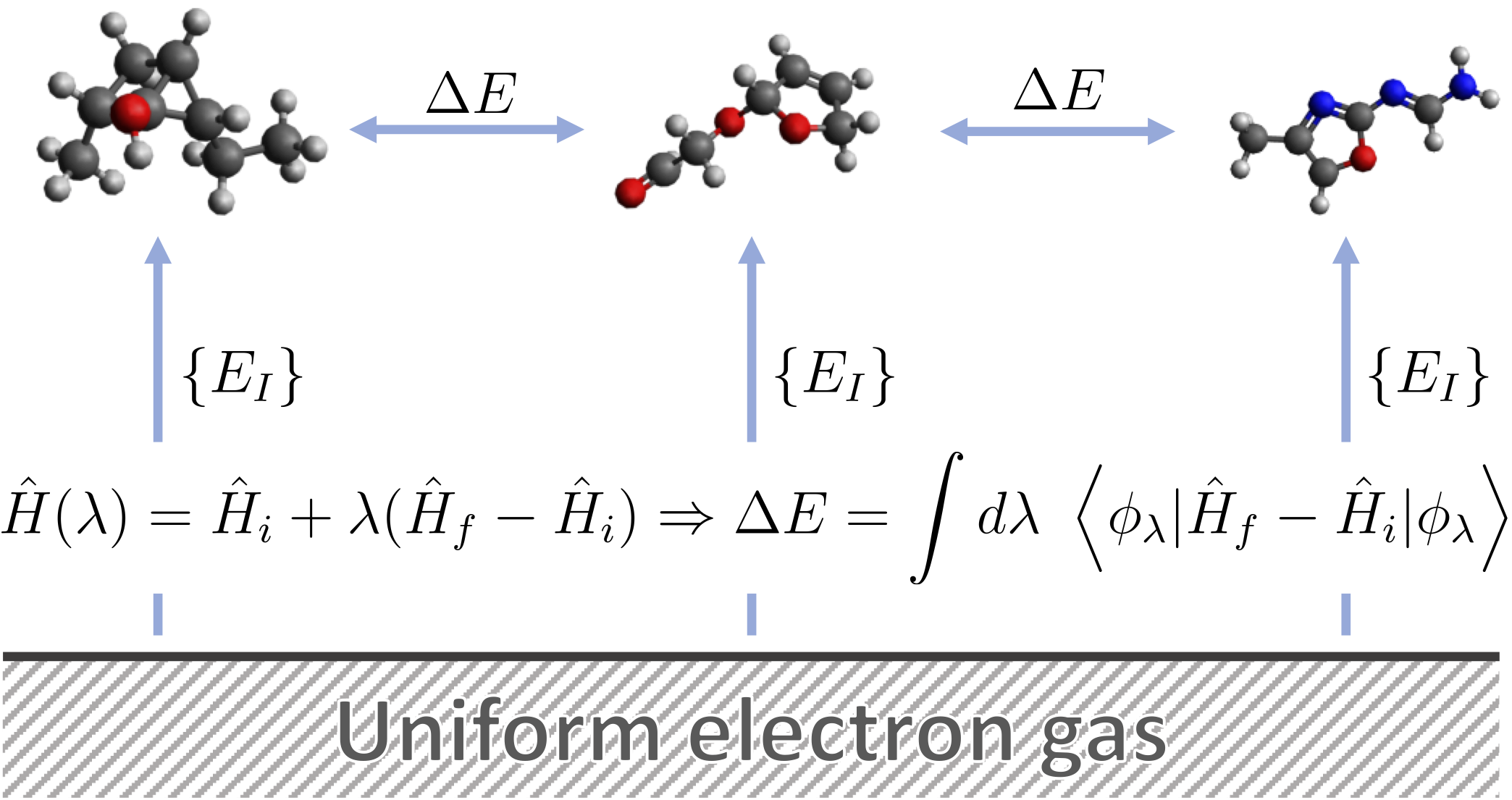}
\caption{Schematic: Starting from the uniform electron gas of appropriate number of electrons, 
thermodynamic integration over linearly grown nuclei enables the unambiguous assignment
of atomic energy contributions enabling direct comparisons between molecules.}
\label{fig:ueg}
\end{figure}

Insertion of $\mu_I$ in Eq.~\ref{eq:alchemy} results in
\bea
\Delta E & = & \sum_I \Delta Z_I \int_0^1 d\lambda \int d\fatr \, \frac{\rho_\lambda(\fatr)}{|\fatr-\fatR_I|} \nonumber\\
& = &  \sum_I \Delta Z_I \int d\fatr \frac{\tilde{\rho}(\fatr)}{|\fatr-\fatR_I|} \; \equiv \; \sum_I \tilde{\mu}_I \Delta Z_I \nonumber\\
& = & \sum_I \Delta E_I \label{eq:AtomicEnergy}
\eea 
an expression consistent with the molecular grand-canonical ensemble DFT~\cite{anatole-jcp2006-2}, and
where $\tilde{\rho}(\fatr) =  \int d\lambda \, \rho_\lambda(\fatr)$, the $\lambda$-averaged density for the path of interest, introduced within APDFT~\cite{vonRudorff2018apdft}, and where $\tilde{\mu}_I$ represents the corresponding averaged alchemical potential. 
The product of $\tilde{\mu}_I$ and its associated change in nuclear charge results in a natural definition of the change in the atomic energy due to the coupling of any initial reference state to any final target state. 
Since also the electron density is a state function, the expression for its atomic decomposition is also obtained via chain-rule in complete analogy to above,
\bea
\Delta \rho(\fatr) & = & \int_0^1 d\lambda \left.\frac{\partial \rho(\fatr)}{\partial \lambda} \right|_\lambda \longeq \sum_I \Delta Z_I \int_0^1 d\lambda \left. \frac{\partial \rho(\fatr)}{\partial Z_I}\right|_\lambda \nonumber\\
& \equiv & \sum_I \tilde{\rho}'_I(\fatr) \Delta Z_I = \sum_I \rho_I(\fatr) \label{eq:rho}
\eea

Note that the choice of reference state and the linearity of path  are arbitrary, and that there is no constraint on the respective atomic contributions other than that their sum must result in the molecular energy. 
As such, atomic contributions are path functions, and cannot unequivocally be determined.  
At least the question of the ideal reference could be plausibly addressed by arguing that geometrically and compositionally highly symmetric systems, e.g.~homo-nuclear molecules such as N$_2$, could be preferable since their atomic contributions to the electronic energy of an atom in its environment are trivially accessible by mere division of the molecular energy by number of nuclei. The total potential energy of an atom in any other molecule can subsequently be obtained by virtue of Eq.~(\ref{eq:AtomicEnergy}) followed by addition of the corresponding atomic nuclear-nuclear repulsion contribution,
\begin{align}
    E_\text{NN} = \sum_I\frac{Z_I}{2}\sum_{J\neq I}\frac{Z_J}{|\mathbf{R}_{I}-\fatR_J|} = \sum_I E_{{\rm NN},I}\label{eq:enn}
\end{align}
However, such a choice, as plausible as it may be, is still arbitrary, and might not be equally well suited for all target systems ({\em vide infra}).

Here, we propose to use the limit of no composition, i.e.~the iso-electronic uniform electron gas (UEG), aka.~jellium, as the most general reference system for CCS instead. 
Any compound can then be generated {\em ex nihilo} by simply scaling up the nuclear charges at the target's geometry in a linear fashion. 
Since this procedure relies on the exact {\em same} reference and path for {\em any} arbitrary target compound,
a unique definition of absolute electronic atomic energy and electron density contributions of AIMs is guaranteed. 
These APDFT based AIMs therefore enable a meaningful 
and comparative discussion of trends of these atomic properties throughout CCS. This idea is illustrated in Fig.~\ref{fig:ueg}. 
Other partitioning approaches, e.g.~Hirshfeld partitioning, rely on the free atom as an intuitive reference. In principle, such a choice, i.e.~the dissociated atom limit, would also be consistent with our framework, e.g.~by annihilating free dissociated atoms while simultaneously growing them back into their molecular framework. The choice of iso-electronic jellium, however, strikes us as preferable since one can expect it to be smoother and more generally applicable. 

\subsection{Computational Details}
Numerically, the $\lambda$-averaged density $\tilde\rho$ is obtained by means of several SCF calculations for fractional nuclear charges $\mathbf{Z_I}(\lambda) = \lambda \mathbf{Z_I}^\text{f} + (1-\lambda)\mathbf{Z_I}^\text{i}$ (using the plane-wave and atomic basis set codes CPMD\cite{cpmd} and HORTON\cite{HORTON}, respectively). For periodic calculations with CPMD, we used a plane-wave cutoff of 200\,Ry and boxes of 15-20\,\AA. Non-periodic calculations in HORTON performed with 6-31G basis set\cite{Hehre1972}. Fractional nuclear charges are encoded by fractional core charges or linearly interpolated GTH\cite{Goedecker1996} pseudopotential parameters for HORTON and CPMD, respectively. To allow for comparison of atomic energy differences with total energies of isolated non-periodic molecules, non-periodic single points energies are calculated at the same level of theory as the corresponding periodic setups, just with the atomic basis set instead of plane-waves. cc-pVDZ/CCSD atomic energies are calculated with MRCC using the \textit{popul=deco} keyword\cite{Kallay2001,mrcc}.

We evaluate the electron density on a either a rectangular or Becke-Lebedev\cite{Lebedev1999,Becke1988} integration grid and use the trapezoidal rule for numerical evaluation of the spatial integral over $\tilde\rho$, obtained with Kohn-Sham density functional theory~\cite{HK,KS} within the local density approximation (LDA)~\cite{ceperley-xc} or Hartree-Fock (HF), as noted where used. 
Density responses are obtained from finite differences with fractional nuclear charges in steps of 0.1$e$. For all APDFT calculations, we verified that the valence electron density and its derivatives $\partial_{Z_I}\rho$ integrate to the total number of valence electrons or zero, respectively.
Note that our scheme is independent of the employed level of theory as long as explicit electron densities for fractional nuclear charges are available.

\section{Results and Discussion}
\subsection{Meaningless atomic energies}
As mentioned before, the choice of both, the reference system, and the path along which the alchemical change is followed, is arbitrary:
Any integrable interpolation which meets the end-points will lead to the correct molecular energy and density. 
This can result in arbitrary atomic energies since these contributions are not state functions (unlike molecular energies), 
i.e.~the resulting energy decomposition becomes path-dependent, and a general comparison across CCS becomes meaningless. 
This is obviously manifested in Eqs.~(\ref{eq:AtomicEnergy},\ref{eq:rho}), where sites without change in nuclear charge between reference and target do not change in their contribution to energy or electron density, respectively. 

\begin{figure}
\centering
\includegraphics[width=.8\columnwidth]{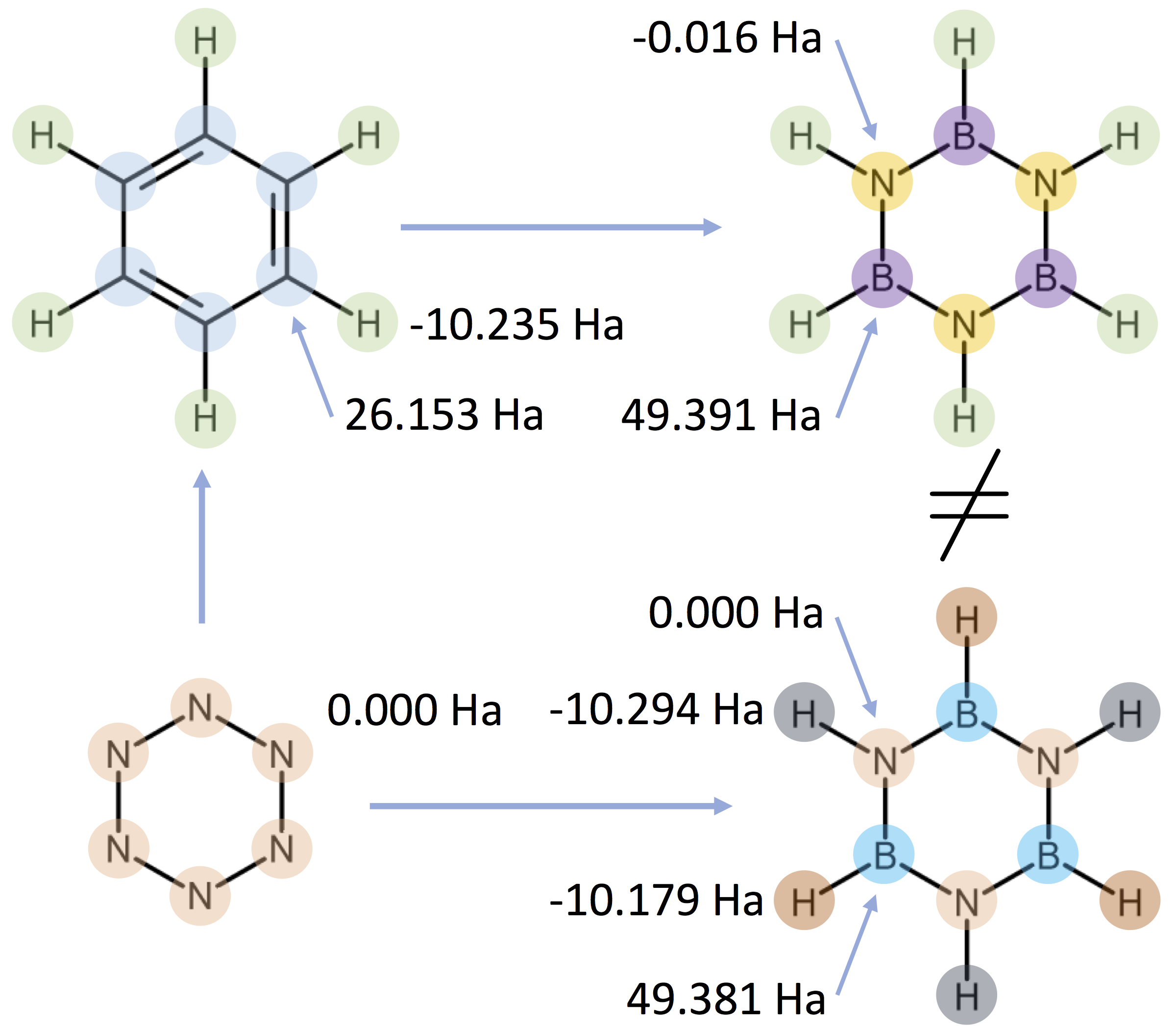}
%\hspace{.5em}\includegraphics[width=.35\columnwidth]{otmr-crop.pdf}
\caption{Path dependent atomic energies of borazine (BNBNBNH$_6$) using Eq.~\ref{eq:AtomicEnergy} and hexazine (N$_6$) as a reference.
Colors encode atomic energies.
All coordinates correspond to the geometry of benzene.
Level of theory is HF/STO-6G.
}
\label{fig:absurd}
\end{figure}

We have collected numerical evidence to illustrate the lack of meaning due 
to  such path dependency in Figure~\ref{fig:absurd}. 
Choosing the direct path from the fully symmetric N$_6$ hexazine ring to fully BN-doped borazine results in three Nitrogen sites with an atomic energy and density identical to the one in N$_6$, implying that for these atoms ``nothing changed''. 
Alternatively, choosing a path via the isoelectronic benzene molecule as an intermediate, however, 
removes this restriction and allows all Nitrogen sites to undergo a change in atomic energy.
Now, however, all hydrogens of borazine are supposed to have the same energy and electron density as in benzene, implying now
that for them ``nothing changed'' (w.r.t. benzene). 
Both end results therefore clearly conflict with the truism that all atoms should contribute to the energy in a molecule. 
Furthermore, the absurdity of these borazine results, obtained by naively following two legitimate, yet different, paths 
becomes obvious when noting the discrepancies among the predictions for the same atomic nuclei:
Energy contributions of Nitrogen, Boron, and Hydrogen atoms differ by 0.016, 0.010, and $\sim$0.057 Ha, respectively. 
If such ``uncertainty'' can already be obtained by simply following reasonable paths it is clear that one cannot rely on such naive decomposition schemes to provide meaningful information for ensuing analysis. 
In consequence, this inconsistency results in an unphysical energy buildup on sites for multiple consecutive rounds of the paths in Figure~\ref{fig:absurd}. 
To ensure that the energy decomposition is well-defined, a unique definition of both, reference system and alchemical path, is required.

To allow for a common reference system for all of CCS, we use the aforementioned isoelectronic UEG as reference, where the alchemical path is given by linear growth of the nuclear charge\cite{Wilson1962}, as shown in Figure~\ref{fig:ueg}. This way, all isoelectronic molecules connect to one commonly shared reference that is of the highest possible symmetry. The advantage is that all reference systems can in principal be enumerated and are shared for all isoelectronic molecules. Similar to molecular orbitals of which there are arbitrary many distinct sets that satisfy the Kohn-Sham equations, each choice of reference system and alchemical path in itself is arbitrary. However, the resulting energies are valid metrics for the atomic contributions to molecular stability via the electronic energy, similar to the way maximally localized orbitals or atomic charges are inspection tools without corresponding experimental observable. 

\begin{figure}
\centering
\includegraphics[width=\columnwidth]{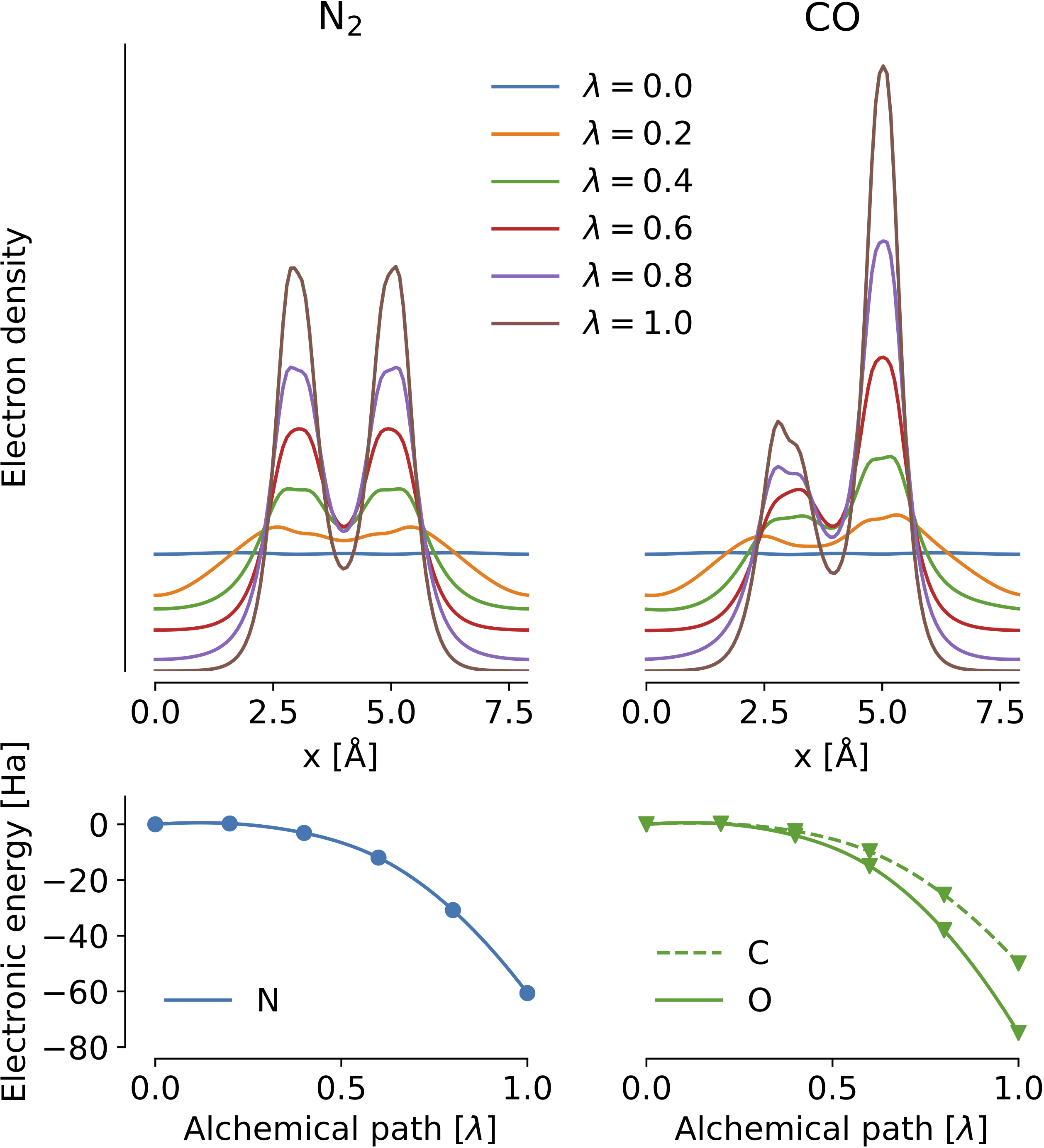}
\caption{Top: $zy$ integrated electron densities as a function of $x$ for N$_2$ and CO and various interpolation parameter values of $\lambda$. 
Going from $\lambda=0$ (uniform electron gas) to $\lambda=1$ (the target molecule) gives self-consistent electron densities for fractional nuclear charges.
Bottom: Build-up of the atomic  contributions to the molecular electronic energy (without nuclear repulsion) according to Eq.~(\ref{eq:AtomicEnergy}
as a function of integration limit, i.e.~integration over $\tilde\rho$ only up to a certain $\lambda$.
}
\label{fig:densgrow}
\end{figure}

\subsection{Diatomics}
As a simple example, we separate the atomic contributions to the electronic energy, i.e. the total energy without the nuclear-nuclear interaction, for neutral diatomic molecules of 14 protons. For the alchemical growth of N$_2$ and CO, Figure~\ref{fig:densgrow} shows the intermediate electron densities for various values of coupling parameter $\lambda$. Starting from the constant electron density of the UEG as reference, the increasing fractional nuclear charges accrue more and more electron density around the nuclei. At the beginning of the alchemical path, the change in density is minute and becomes larger towards the end of the path. 
A zero gradient at $\lambda = 0$ is to be expected due to the maximal symmetry of the electron density, as also noted previously~\cite{Fias2018}.
The $\lambda$-integral of the intermediate electron densities constitute $\tilde\rho$ in eq.~\ref{eq:AtomicEnergy}. 
The nonlinear behavior in density is 
also reflected in the corresponding energies  in Figure~\ref{fig:densgrow}, where the last 40\% of the alchemical path account for $\sim$75\% of the energy change.
While the atomic energy of N trivially reaches half the energy of N$_2$, the decomposition of the molecular electronic energy of CO indicates that oxygen contributes (substantially) more to the molecular electronic energy  than carbon. This is not surprising since, neglecting chemical bonding, the electronic energy scales with nuclear charges as $E\sim -0.5 \sum_I Z_I^{7/3}$. 

\begin{figure}
\centering
\includegraphics[width=\columnwidth]{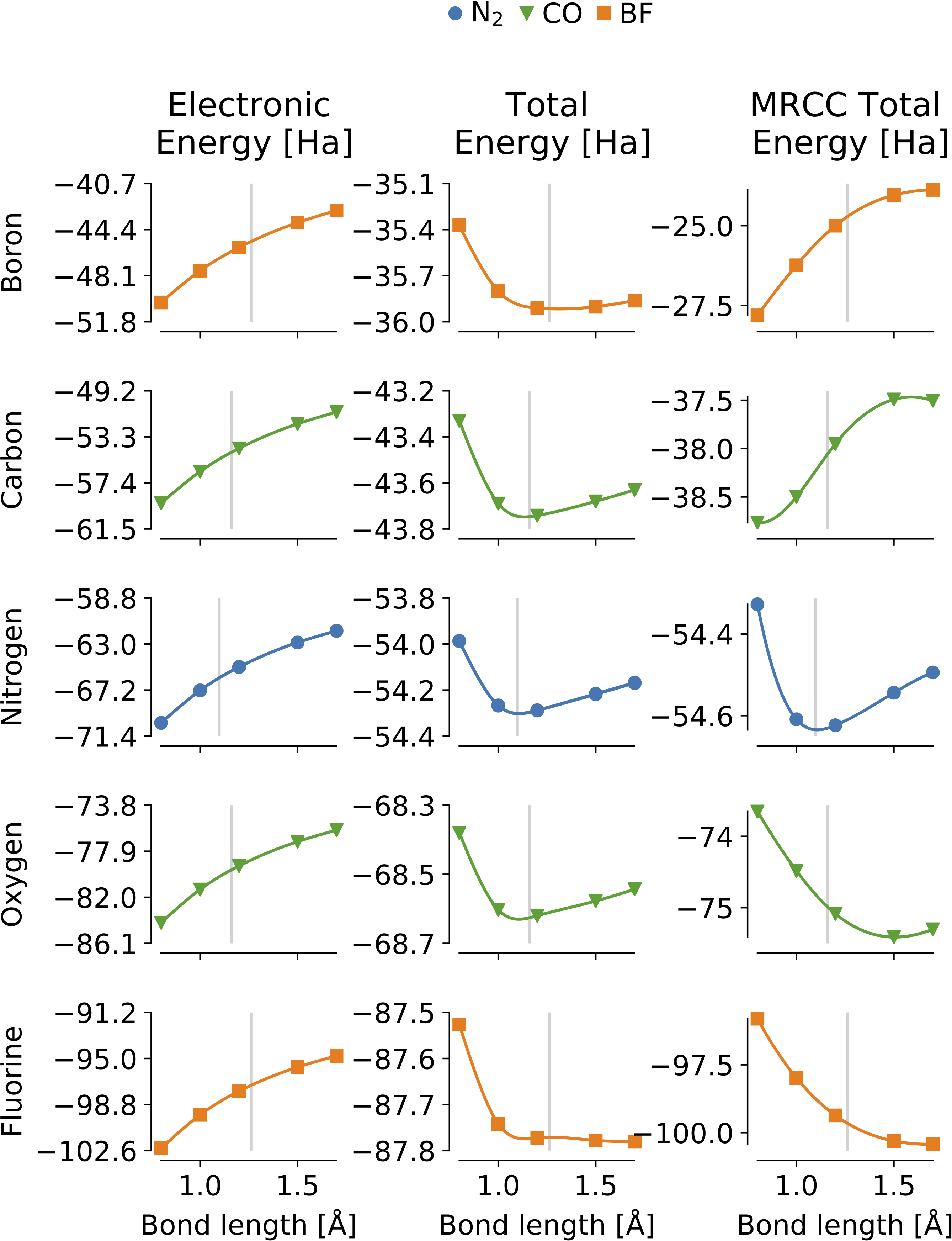}
\caption{Atomic contributions to the electronic (left) and total (center) energy as a function of nuclear distance $d$ in neutral iso-electronic molecules N$_2$ (blue circles), CO (green triangles), and BF (orange squares). For comparison, atomic contributions to total cc-pVDZ/CCSD energies are shown (right).
%The small wiggle for F in the total energy of BF is due to a numerical instability.
Vertical lines denote experimental equilibrium distances.
}
\label{fig:dimer}
\end{figure}

Figure~\ref{fig:dimer} shows the atomic energy contributions for different nuclear distances in N$_2$, CO, and BF. For large separations, the energy grows towards the free atom limit, strictly required for any valid atomic energy partitioning scheme. For short bond distances, all atoms gain energy due to interaction with the more compact electron density. Again, we confirm that atoms with larger nuclear charges contribute more to the electronic energy. 
Since the total energy also contains the nuclear-nuclear interaction energy (which for diatomic molecules is split evenly across nuclei as per Eq.~(\ref{eq:enn})), the repulsive wall becomes dominant for small interatomic distances, while the free atom energy is still recovered in the limit of infinite interatomic distances. Figure~\ref{fig:dimer} shows the distance dependency of the atomic contributions with nuclear-nuclear interactions. The sum of the two atomic contributions to the total energy recovers the well-known potential energy surface of the diatomic molecules. 

It is interesting to compare these results to the cc-pVDZ/CCSD atomic energies from MRCC, also shown in Figure~\ref{fig:dimer}. Their atomic contributions are qualitatively different: The density matrix based CCSD energy expression is partitioned according to the atomic basis functions. At equilibrium distance, this results in an attractive force for only the lighter atom of the dimer, while the heavier atom is being repelled. Compensation of the two effects leads to total energy minimum. This is effect is much larger than the difference between the CCSD and LDA potential energy surface which can be assessed by comparison of the (symmetric) Nitrogen atomic contributions for the two methods. From the APDFT point of view, the CCSD partitioning represents a different reference (free atoms) and path (non-linear increasing overlap of atom centered basis functions). 

\subsection{Various molecules}

\begin{figure}
\centering
\includegraphics[width=\columnwidth]{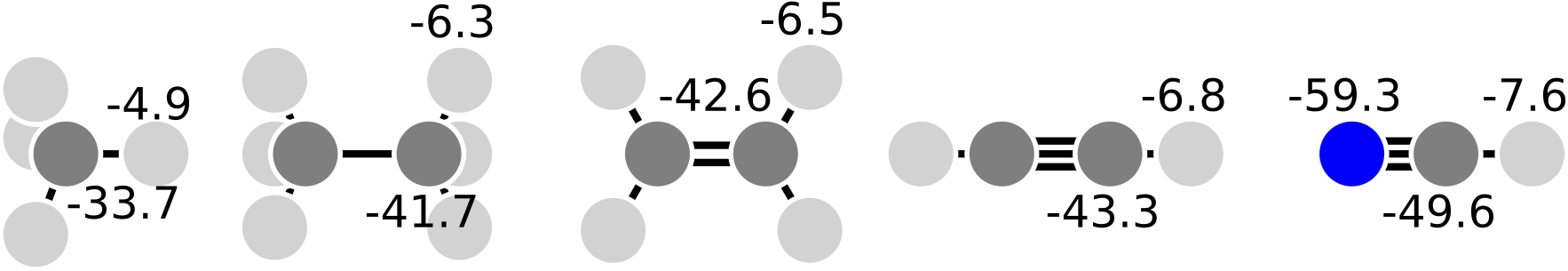}
\vspace{1em}
\rule{\columnwidth}{0.01cm} \newline
\includegraphics[width=\columnwidth]{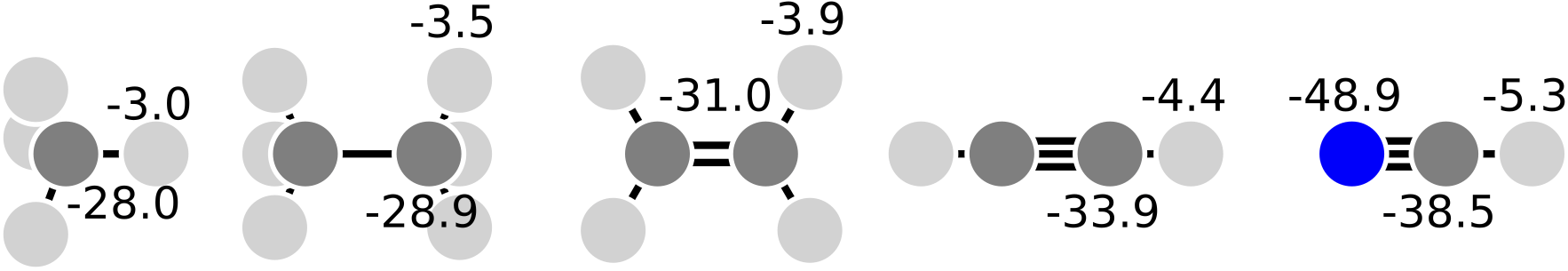}
\caption{Atomic contributions to the electronic (top) and total (bottom) energy [Ha] in various molecules. 
From left to right: methane ($sp^3$), ethane ($sp^3$), ethylene ($sp^2$), acetylene ($sp$), and hydrogen cyanide ($sp$). 
Identical energies for symmetry-equivalent atoms omitted.
Subtraction of corresponding free atom energy values for LDA/6-31G
(-0.48, -37.37, and -54.11 Ha for H, C, and N, respectively) enables atomization energy estimates.}
\label{fig:spn}
\end{figure}

Due to their chemical diversity the atoms in the molecules on display in Fig.~\ref{fig:spn} may be of interest. Atomic energy contributions to the electronic and total molecular energy is shown, as well as free atom's energies. Subtraction of the latter from the former two enables direct estimates of the corresponding contributions to covalent binding.
It is interesting to note that even in the electronic energy case, the energy of the carbon atom in methane is less than for the free atom.
The hydrogen atoms, conversely, overcompensate by being stabilized by multiple Hartrees. 
The scale and large variance among these results suggests that their interpretation has to be carried out with great caution. 

When comparing the atomic energies among the different molecules, however, plausible trends emerge: The carbon atom is increasingly stabilized as it participates in bonds with atoms that have increasingly more electrons.
For all hydro-carbons (i.e. with the exception of hydrogen cyanide), however, carbon's atomic contribution to the total atomization energy
remains repulsive. In hydrogen cyanide, it is again the heavier atom, nitrogen, which is repelled, while hydrogen and now also carbon overcompensate. 
These results imply an interpretation of the hetero-nuclear covalent bond where the  atom with the larger nuclear charge has nothing to ``gain'' from forming bonds but is rather being ``pulled'' in by earlier elements which are stabilized by the additional electrons.

In order to compare APDFT AIM results for these molecules also 
to other methods, we report atomic contributions to the atomization energy for all molecules in Fig.~\ref{fig:spn} in Tab.~\ref{tab:atomicenergies}, along with 
cc-pVDZ/HF and CCSD numbers from MRCC, as well as PhysNet\cite{PhysNet2019} numbers. 
PhysNet is the revised implementation of the neural network 
introduced in Ref.~\cite{unke2018reactive}, and has been
trained on over 110'000 organic molecules from the QM9 data set~\cite{QM9}. 
Within the view of APDFT, the PhysNet numbers should correspond to a choice of reference and path as atomization energy training set mean and some complex non-linear regressed interpolation path, respectively.
We note that the scale of the variance among all atomic contributions decreases dramatically as one goes from APDFT (Ha's) to MRCC (100s of mHa) to PhysNet (mHa's).  Furthermore, positive contributions are common among the heavier atoms within APDFT, while they are rare for MRCC, and inexistent for PhysNet. 

More specifically, within APDFT, the atomic  contributions of hydrogen and carbon to the atomization energy increase systematically when going from methane to ethane to ethylene to acetylene, and to hydrogen cyanide. 
For the latter, carbon changes from repulsive to attractive (due to the presence of nitrogen which contributes a large positive amount). 
By contrast, MRCC's atomic contribution are all mostly attractive (with the exception of hydrogen in HCN) within CCSD, 
and within HF (with the exceptiong of hydrogen in HCN and carbon in acetylene).
The atomic electron correlation contribution to the atomization energy is estimated by the difference between the latter,
and always strengthens the binding for carbon. In the case of hydrogen, the trend is less clear; 
and for nitrogen in HCN the electron correlation energy weakens the binding.
Comparing the atomic contributions across different molecules, it is interesting to note that the trend for CCSD and HF seems to be reversed with respect to APDFT for the hydrogen atom: The atomic energy contribution to the atomization energy decreases as one goes from methane, to ethylene, to hydrogen cyanide, the only notable exception being acetylene with an unusually large attractive contribution. 
For carbon, CCSD and HF do not exhibit clear trends among the few molecules discussed. They do, however, 
complement the respectively small or large (in the case of acetylene) amount provided by the hydrogen atoms:
Most interestingly, in acetylene, carbon hardly contributes at all to the atomization energy according to CCSD. 
When considering the atomic contributions according to PhysNet, the variation among the molecules for given elements
is often on the scale of mHa. For hydrogen, the same systematic increase in binding is observed as for APDFT, 
except for the case of HCN, where the hydrogen strengthens binding less than in ethane. 
For carbon, the trend is less obvious: It contributes more in methane than in ethane. 
But starting from ethane a systematic increase by 9 and 3 mHa is observed
when reducing the single to double and triple bond, respectively. Again, carbon in HCN
breaks the trend by contributing less than in ethylene. 
These results demonstrate, yet again, how using different partitioning schemes leads not only to 
different estimates but also to different trends within molecules as well as across chemical space.

\begin{table}[]
    \begin{ruledtabular}
    \begin{tabular}{llrrrr}
&&APDFT&CCSD&HF&PhysNet\\\hline
H&Methane&	-2.53	&-0.089	&-0.089	&-0.109\\
&Ethane	&-3.03&	-0.074&	-0.077&	-0.114\\
&Ethylene&	-3.43&	-0.087	&-0.065	&-0.114\\
&Acetylene&	-3.93&	-0.270&	-0.222	&-0.117\\
&HCN&	-4.83&	0.018&	0.076&	-0.112\\\hline
C&Methane&9.37&-0.270&-0.157&-0.198\\
&Ethane&8.47&-0.309&-0.201&-0.195\\
&Ethylene&6.37&-0.239&-0.205&-0.204\\
&Acetylene&3.47&-0.006&0.017&-0.207\\
&HCN&-1.13&-0.358&-0.175&-0.201\\\hline
N&HCN&5.21&-0.100&-0.207&-0.169\\
    \end{tabular}
    \end{ruledtabular}
    \caption{AIM energy contribution estimates to the calculated atomization energy [Ha] for the molecules in Figure~\ref{fig:spn} according to APDFT (LDA level of theory), the basis-function decomposition as implemented in MRCC\cite{Kallay2001,mrcc} at HF/cc-pVDZ and CCSD/cc-pVDZ level of theory, and PhysNet~\cite{PhysNet2019} neural network~\cite{OliverUnke}. All methods evaluated for identical geometries. Symmetry equivalent sites have identical contributions.}
    \label{tab:atomicenergies}
\end{table}{}

\subsection{Atomic electron densities}
\begin{figure}[t!]
\centering
\includegraphics[width=\columnwidth]{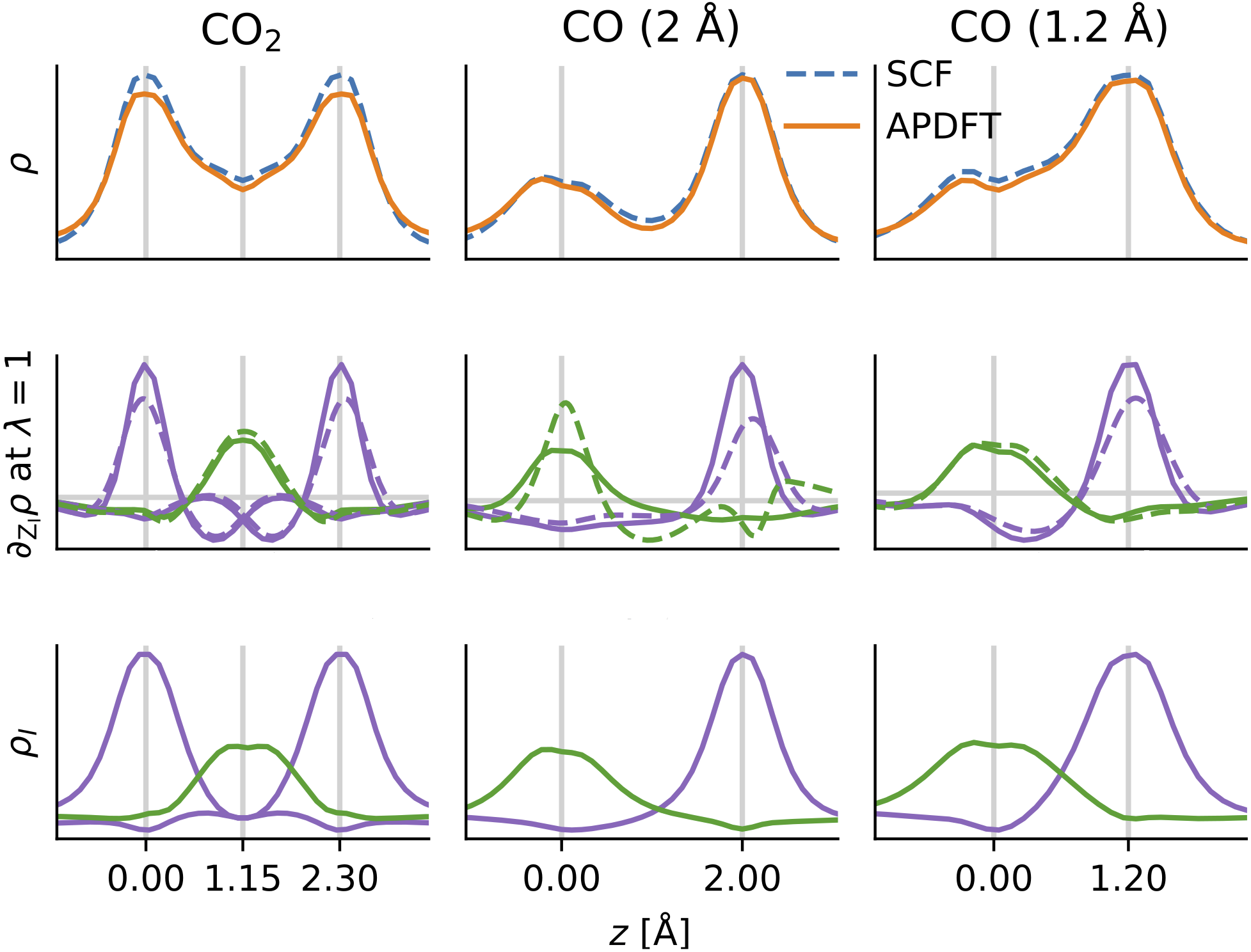}
\caption{$xy$-integrated slices of atomic valence electron densities for CO and CO$_2$ projected onto the bond axis ($z$). 
Vertical gray lines denote nuclear positions. Top: Electron density $\rho$ as obtained from integrating the density responses (APDFT) and from regular CPMD. 
Mid: response of the molecule's electron density to a change in $Z_I$ only, i.e. $\partial_{\rm{Z_I}}\rho|_{\lambda = 1}$. Shown both for plane-wave CPMD (stroked) and -- for comparison -- atom-centered orbital calculations with CP2K and PBE/DZVP-MOLOPT-SR-GTH (dashed).
Bottom: Decomposition the electron density into atomic contributions according to eq.~\ref{eq:rho}. 
%Spatially integrated symmmetrized atomic contributions give the atomic charge [$e-$]. 
Periodic unit cell of 15\,\AA, (CO$_2$) or 20\,\AA,(CO) side length.}
\label{fig:atomicdensities}
\end{figure}

Figure~\ref{fig:atomicdensities} shows electron densities (total, response, and atomic according to eq~\ref{eq:rho})
for CO$_2$ at 1.15 {\AA}, and CO at 2 and 1.2 {\AA} interatomic CO distance. 
Along the alchemical path of growing a molecule from the uniform electron gas, we obtain the responses of the electron density due to changes in nuclear charges. If integrated over the full path, these responses recover the electron density difference between target (the regular molecule) and reference (jellium). The top panel in Figure~\ref{fig:atomicdensities} compares the total electron density as obtained from a regular SCF calculation with the electron density obtained via numerical integration. For all systems and in complete analogy to the atomic energies, summation over atomic electron density contributions recovers the correct (i.e. self-consistent) overall electron density distribution. The electron density distribution is well recovered for both bonded and loosely interacting configurations, as shown for CO at both 1.2\,\AA\, and 2\,\AA\, interatomic separation. 
%Only for the CO$_2$ calculations, a slight asymmetry is visible, which illustrates the effects of the finite unit cell in the fully periodic calculations. 

The electron density responses shown in the mid panel of Figure~\ref{fig:atomicdensities} have their dominating contribution around the atom the nuclear charge of which is being perturbed, which is in line with the expectation that the electron density response should be localized in the vicinity to the perturbation. This effect is visible for both the periodic plane-wave calculations which allow to connect to the UEG and for the equivalent non-periodic setup with atom-centered orbitals. In direct comparison, the two methods give very similar results for the electron density response at $\lambda = 1$, i.e.~a point of the alchemical path which can easily be evaluated with both basis sets. They differ most in the case of extended CO where the atomic basis sets have less overlap between the atomic sites. This is in line with previous observations on the behavior of atomic basis sets when performing alchemical changes far from equilibrium geometries\cite{vonRudorff2018apdft}.

The bottom panel of Figure~\ref{fig:atomicdensities} showcases the partitioned electron densities. 
At another atom, the atomic electron density response is (close to) zero, which can be understood with the help of Kato's theorem\cite{Kato1957} which states that the electron density at a nucleus is proportional to the spherically averaged electron density gradient. The proportionality constant only depends on the nuclear charge of the atom in question. Since the electron density at the nucleus is independent of the chemical environment, this means that in a small volume around an atom the electron density distribution can change if and only if the nuclear charge changes. Therefore, if there was a non-zero electron density contribution of atom $I$ at the nucleus of atom $J$, this contribution would need to decay with distance from nucleus $J$ in order to follow Kato's theorem despite the increased density at the nucleus. This however, would yield two cusps in the electron density of atom $I$: one at the nuclear site of atom $I$ and one at the nuclear site of atom $J$. Contributions to the electron density cusp of one atom coming from other atoms, however, is unphysical since it would violate the well established frozen core assumption which, for example, forms the basis of the pseudo-potential approximation which has been shown to work well in countless studies. 
Therefore, to avoid the  unphysical contribution to another atom's cusp, and to avoid violating Kato's theorem, the charge contribution needs to approach zero close to the nucleus of any other atom. In other charge partitioning schemes, like e.g. Hirshfeld charges, this is not a given. Even though the aim of Hirshfeld charges is to partition the electron densities such that each atomic contribution is as similar as possible to the corresponding free atom, the assignment of part of the electron density at nuclei to other atoms renders the atomic densities at the nuclei strongly dependent on the free atom electron density of the chemical environment.

Note that the requirement that the atomic contributions become zero at other nuclei does not specify how quickly the atomic contribution to the electron density reaches zero. Bader charges would be an extreme case where zero contribution would be reached at the electron density minimum between atoms, while another extreme would be a decay of atomic contribution that reaches zero just before another nuclear site appears. From Figure~\ref{fig:atomicdensities}, it is evident that the latter is the case for atomic densities obtained by the APDFT method---at least for the molecules presented here. This means that the picture of interacting electron densities stabilizing a molecule is still applicable: the resulting density overlap increases for decreasing interatomic distances.

Each electron density response in Figure~\ref{fig:atomicdensities} is charge-neutral, i.e. the spatial integral thereof evaluates to zero---which can be used as a criterion to assess numerical stability of a given setup and convergence in terms of box size and plane-wave cutoff. In order to obtain the atomic electron density contributions, the electron density of the reference system (the uniform electron gas) would need to be distributed over all atomic sites. With the observation that atomic electron densities need to be positive everywhere and zero only close to the nuclei, this distribution is uniquely defined. Away from the nuclei, all $\lambda$-integrated atomic density responses sum up to the negative of the reference density, since all electrons have moved towards the nuclei. Each atomic density response is therefore assigned as much charge of the reference electron density as is necessary to shift the atomic contribution to the electron density to zero. This is not a degree of freedom, but rather the only way how AIMs in APDFT avoid unphysical negative charge build-up.

%Integrating the atomic electron densities yields atomic charges close to the valence electron count, and consistent with the atomic %energies shown in Table~\ref{tab:atomicenergies}: 
%The heavier atom looses a bit of electron density, in line with the heavier atom loosing energy w.r.t. the free atom in the discussion of atomization energies above. 
%Interestingly, the integrated charge densities happen to be close to valence electron numbers, even though there is no such requirement in our method. 
%This supports the choice of reference system and alchemical path to give reasonable results. 

\begin{figure*}[t!]
\centering
\includegraphics[width=\textwidth]{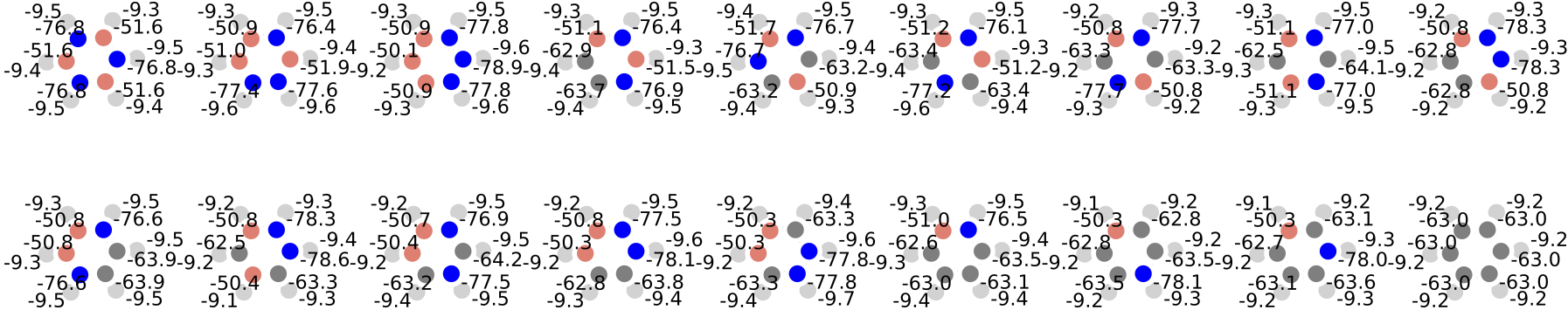}\\
\vspace{1em}
\rule{\textwidth}{0.01cm} \newline 
\vspace{1em}
\includegraphics[width=\textwidth]{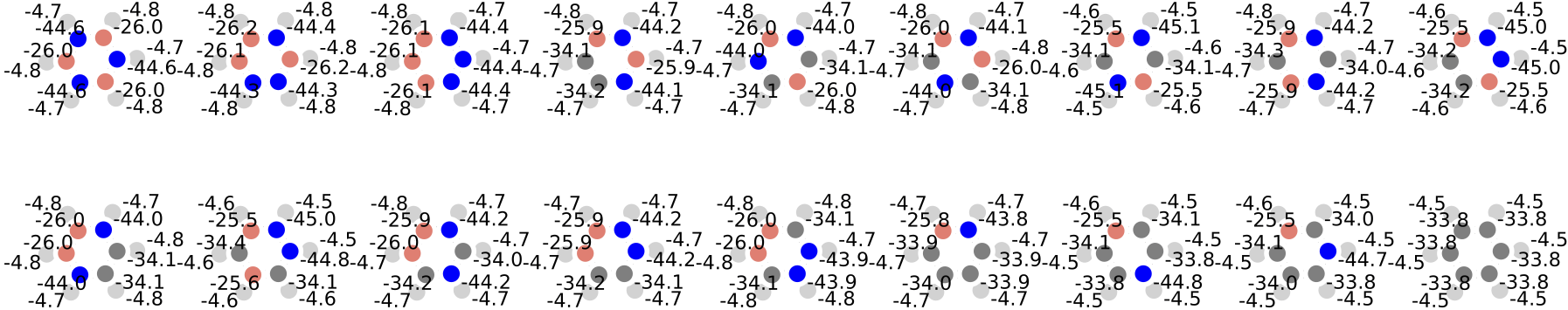}
\caption{Atomic contributions to the total energy (bottom two rows) and electronic energy (top two rows) in Ha for all BN-doped benzenes. All molecules sorted by their total energy.
Atomization energies can be recovered by subtraction of 
free atom energies [Ha] at LDA/6-31G level of theory for
H: -0.48,
B: -24.34,
C: -37.37, and
N: -54.11
}
\label{fig:c6bn}
\end{figure*}

\subsection{BN doped benzene derivatives}
In order to broaden the analysis of APDFT based atomic energies, we report atomic energies for all the iso-electronic 18 unique benzene derivatives in the geometry of benzene, obtained by doping with BN pairs. Molecular energetics of such systems have already been explored with alchemy previously~\cite{anatole-jcp2007,CCSexploration_balawender2013,Fias2018}.
Again, we find that the heavier atoms are ``repelled'' from the binding, and that this effect is overcompensated by elements with smaller nuclear charge, hydrogen in particular. 

The results in Figure~\ref{fig:c6bn} shows a clear separation of the atomic energy contributions by element, since the absolute energy values are dominated by free atom energies. However, the variance among the same elements as a function of their environments is quite pronounced. Considering electronic energy alone, the typical spread between minimal and maximal atomic contributions is about 1.25\,Ha, whereas the total molecular energy varies by 0.6\,Ha on average. On the energy scale relevant for chemistry, this showcases a strong sensitivity towards the local chemical environment. 

It is worth noting that this spread in energy contributions also affects distinct groups of atoms within the same molecule. As shown in Figure~\ref{fig:c6bn}, the NH groups in BNBNNBH$_6$ have the largest difference (0.2\,Ha) in electronic energy contributions of any otherwise identical groups in our data set. For all sites that are identical due to symmetry, however, there is no difference in atomic energy contributions. This represents an important consistency check, since atoms in the same local environment should have the same atomic energy contributions.

%Comparing the total molecular energies for all nuclear configurations would only allow to quantify the energy contribution of a BN pair together since this group denotes the smallest common change between molecules. Furthermore, the estimation of the BN group energy e.g. by means of regularized linear regression would yield a value that is non-specific to the particular molecular geometry and therefore, subject to the aforementioned spread in total energy contributions.

\begin{figure*}
\centering
\includegraphics[width=\textwidth]{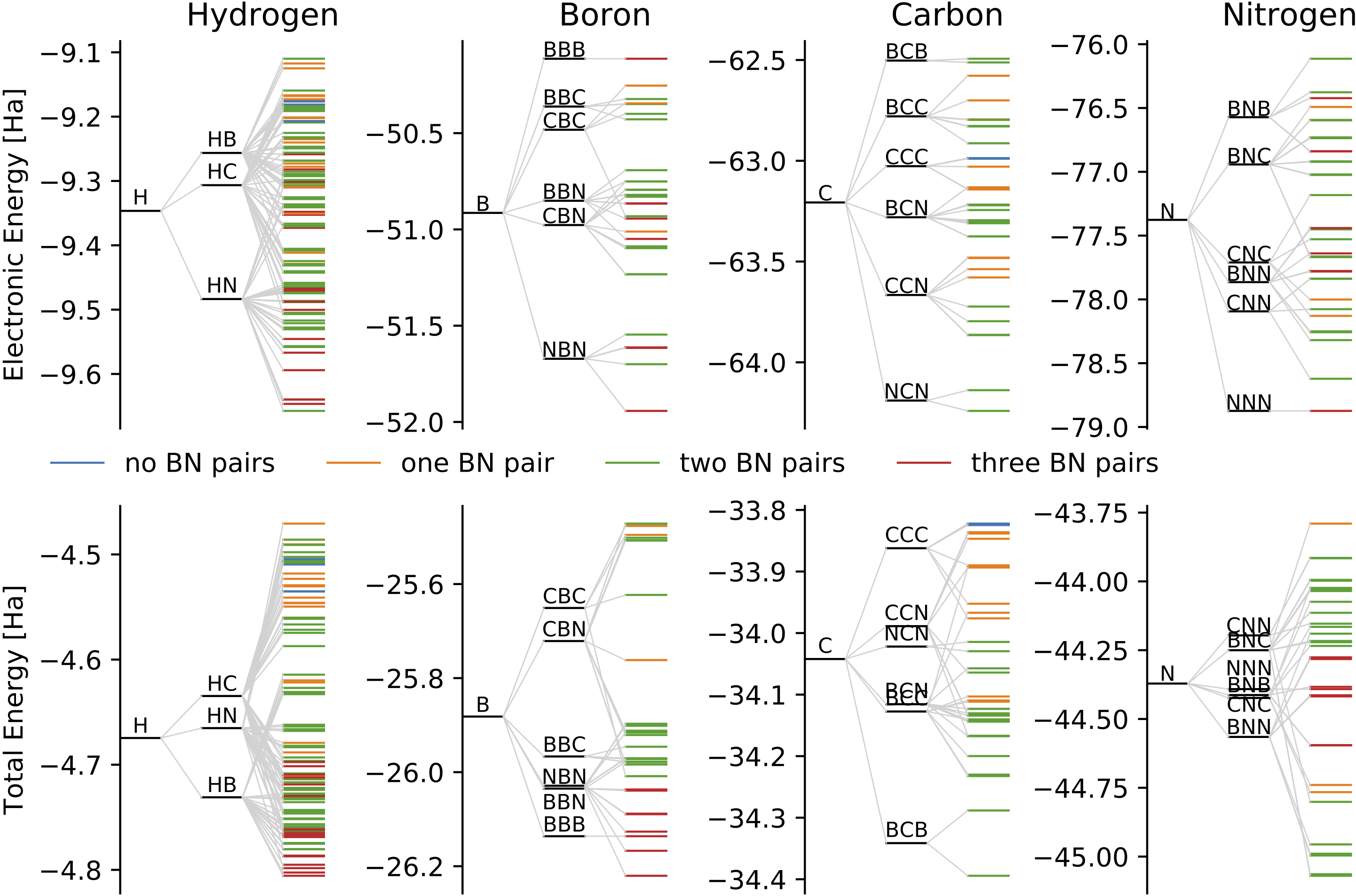}
\caption{Shift in atomic electronic energy (top) and total energy (bottom) due to the chemical environment. For every element, the mean atomic energy is given first. Upon bonding, the energy is shifted (not necessarily lowered), depending on the nearest neighbors. The energies in the individual molecules from Figure~\ref{fig:c6bn} are colored by the number of BN pairs.}
\label{fig:statsplot}
\end{figure*}

To further investigate the impact of the local chemical environment on the atomic energy contributions, Figure~\ref{fig:statsplot} also shows the change in atomic total energy upon embedding an atom in a chemical environment. The energy difference between the average atomic energy and the individual contribution is on the order of about 1\,Ha. This emphasizes that our atomic energies are not independent of chemical environment, i.e. we obtain an energy decomposition from a molecule and its total energy, not building total energies from atomic contributions.

It is interesting to see that the atomic total energies emerge in groups depending on their molecular environment. At no point in the derivation or the implementation, this has been enforced. The groups of chemical environments are clearly visible on the energy scale in Figure~\ref{fig:statsplot}. Moreover, this grouping clearly matches the expectation that each molecular environment is key for the atomic energy contribution. This connects to the force field picture, where different atom types are used to model the atom behavior in different environments. Similarly, in machine learning context, the local environment is often taken as a representation of atomic similarity\cite{Montavon2013} or used to expand total energies in atoms together with their chemical environment, e.g.~within the ``amon'' based framework\cite{Huang2017}. In Figure~\ref{fig:statsplot} it is clearly visible that nearest neighbor sites have the highest impact on atomic total energies, since the splitting in energies between environments defined by the neighbors alone is significantly larger than the splitting between energies from environments if the whole molecule is taken as such. This is in line with the concept of near-sightedness of matter and electronic structure, which is regularly used for linear-scaling or local approximation schemes in quantum chemistry~\cite{StijnPNAS2017}. With our atomic total energy decomposition, it becomes clear that the atomic energies are largely independent of total stoichiometry in larger molecules, 
since the number of BN pairs in Figure~\ref{fig:statsplot} drives no general trend in the splitting.

The ranking of local environments in terms of their average atomic energy contribution is very consistent: the ordering of the mean electronic energies in an environment in Figure~\ref{fig:statsplot} is inverse for nitrogen atoms and for boron atoms. It is important to emphasize, that while the energy expression eq.~\ref{eq:AtomicEnergy} follows a Coulombic expression, this is not a purely electrostatic finding, since the integrated density response $\tilde\rho$ is neither reference nor target density, but rather captures all electronic responses along the alchemical transformation path.

In practice, the accuracy of our approach can be reliably assessed by summing up all atomic contributions and comparing the result to the difference in electronic total energies of the two systems in question. Since our method is exact, remaining deviation comes from both the number of intermediate points considered for evaluating the integral over $\lambda$ and from the numerical integration grid. Both of these potential sources of numerical imprecision can be systematically improved with well-tested established methods. This accuracy estimation requires no additional calculations, since the relevant single point calculations of reference and target compound are part of the alchemical integration path. In our experience, five intermediate points give sufficient accuracy for $\tilde\rho$.

%Cusps at R_I and R_J
%Electron number of atoms? 
%

\section{Conclusion}
%Mention that all properties could have been estimated from some nth order perturbation of the UFE
Within APDFT~\cite{vonRudorff2018apdft}, total atomic energies of atoms in molecule have been calculated 
using thermodynamic integration and application of the chain-rule. 
APDFT treats atomic energies like path functions, making their calculating arbitrary. 
Referencing to the uniform electron gas, however, and relying exclusively linear scaling up of nuclear charges at the desired target compound's geometry 
results in these properties being well-defined in an unambiguous, unique fashion. 
The proposed decomposition of total energies into atomic contributions is very general as it is applicable to any computational chemistry method that 
allows to obtain molecular electron densities for fractional nuclear charges. The method is easy to implement as only two integrals need to be evaluated: 
The one-dimensional integral over $\lambda$ and the three-dimensional integral for $\tilde\rho$ -- both problems have been solved in typical quantum-chemistry codes. 
Our results suggest that the atomic energy contribution is extremely sensitive. 
In the case of the atomization energy it can be of the order of 1\,Ha, even for identical elements in similar molecules. 
Atomic atomization energy contributions can also assume positive and negative values, depending on their relative nuclear charge with respect to the other
atoms present in the molecule. 

With free atom energies commonly used as reference in more established AIM schemes, the APDFT  offers a new way to access atomic energies and electron densities
that are not directly observable, but nevertheless allow for a consistent interpretation of the energetics and electronic structure of matter projected onto contributions from
individual atomic sites. Due to its rooting in first principles, and, more importantly, the need for the UFE as a reference, 
this approach should also be particularly well applicable towards the partitioning of condensed phase systems. 
This scheme might prove useful for future molecular/materials design attempts, for navigating chemical compound space in a more transparent manner,
and, last but not least, to also virtually building up desirable matter one atom at a time.

\begin{acknowledgments}
Both authors thank O.~Unke for generating specific atomic energy estimates for the sake of comparison using PhysNet~\cite{PhysNet2019}, and M.~Kallay for providing the atomic CCSD version of MRCC\cite{Kallay2001,mrcc}.
OAvL is thankful for discussions with S.~Fias. 
We acknowledge support by the Swiss National Science foundation (No.~200021\_175747, NCCR MARVEL).
Some calculations were performed at sciCORE (http://scicore.unibas.ch/) scientific computing core facility at University of Basel.
\end{acknowledgments}

\bibliography{literatur,main}
\end{document}